\documentstyle{article}
\def\half{\frac{1}{2}}
\def\ad{\dot{a}}

\def\be{\begin{equation}}
\def\bea{\begin{eqnarray}}

\def\grad{\nabla}
\def\ee{\end{equation}}
\def\eea{\end{eqnarray}}

\def\Bt{{\cal T}}
\def\dDpx{{d^{D+1}x}}
\def\dDx{{d^D x}}
\def\dDmx{{d^{D-1}x}}

\def\dir{ \partial_r }
\def\dirr{ \partial_{rr} }

\def\wave{ \vcenter{\vbox{\hrule height.3pt \hbox{\vrule width.3pt height9pt
\kern9pt \vrule width.3pt} \hrule height.3pt}} \, }
\def\comma{ \hspace{2mm}, }
\def\period{ \hspace{2mm}. }
\def\s{ \hspace{1mm} }
\def\b{ \hspace{2mm} }

\def\EXP{ {\rm e} }

\def\de{ {\rm d} }
\def\half{ \frac{1}{2} }

\def\minus{ \mbox - \, }

\def\div{ \partial_v }

\newcount\sectionnumber
\def\sect{\global\equationnumber=0
\global\advance\sectionnumber by 1
\the\sectionnumber . }
\newcount\equationnumber
\def   \num
{\eqno{\global\advance\equationnumber by 1
\left(\the\sectionnumber .\the\equationnumber \right)}}

\begin{document}
\title{Lower Dimensional Black Holes: Inside and Out}
\author{R.B. Mann \\
Department of Physics \\
University of Waterloo \\
Waterloo, Ontario \\
N2L 3G1}

\date{January 15, 1995}
\maketitle
\vspace{15pt}

\begin{center}
{\bf Abstract}
\end{center}
\vskip 10 pt

I survey the physics of black holes in two and three spacetime
dimensions, with special attention given to an understanding of their
exterior and interior properties.

\vskip 15 pt

\section{Introduction}

The trickle of interest in lower-dimensional gravity that began a little more
than a
decade ago has turned in recent years into a virtual flood. There are, I think,
two
(related) reasons for this. First, the technical difficulties present in a
wide range of problems in $(3+1)$ dimensional gravitation become significantly
simpler
in lower dimensions. The pedagogical value of this fact was noted some time ago
by Collas
\cite{Collas}: one hopes that lower dimensional gravitational
physics can provide insight into
problems in $(3+1)$ dimensions by yielding a greater measure of computational
simplicity
without sacrificing too much of the conceptual complexity of the original
problem.
A number of physical problems have been approached in this manner, including
 the study of quantum gravitational effects in a mathematically tractable
setting
\cite{Odintsov}, and clarification of the conceptual issues associated with
black hole physics \cite{RGRG,Harvrev}. Second, there are some
physical
systems that are
effectively confined to move in lower dimensions, such as
cosmic strings and domain
walls \cite{cstring}. Indeed, the physics of strings and
membranes necessitates the
introduction of effective  lower-dimensional geometries,
and  understanding of a number of
problems in string theory (motivated by the original work of Polyakov
\cite{Polya}) has been advanced by a study of lower-dimensional gravity.

It is not possible to give a comprehensive review of
lower-dimensional black holes (let alone lower-dimensional gravity),
in the space I have been alloted. Consequently this review
will be somewhat idiosyncratic, expressing my own perspective on the
subject and highlighting a number of issues and viewpoints on lower-dimensional
black
holes that I hope will provide a starting point for those interested in
research
in this field. I apologize in advance to those authors who might feel that
their work
is under-represented here.

I shall begin with a short review of the main features of lower-dimensional
gravity. This
will be followed by a discussion of the physics outside of lower-dimensional
black holes,
including gravitational collapse, mass/energy, thermo\-dynamics, and quantum
properties. I shall then turn to a consideration of the physics inside the
event horizon,
discussing recent research in this subject.

\section{Ins and Outs of Lower-Dimensional Gravity}

Perhaps the most direct way of approaching lower-dimensional gravity is to
begin with
a consideration of Einstein's equations in $(D+1)$ dimensions
\be
G_{\mu\nu} = 8\pi G_{D+1} T_{\mu\nu}  \label{2.1}
\ee
where $G_{D+1}$ is Newton's constant in $(D+1)$-dimensions and and $T_{\mu\nu}$
is the stress-energy tensor.  For $D\ge 3$, a vanishing stress-energy implies a
vanishing Einstein tensor $G_{\mu\nu}$, but not necessarily a vanishing Riemann
tensor
$R_{\mu\nu\alpha\beta}$: it is possible to
have nonzero curvature in regions of spacetime where there is no stress-energy.

This feature is lost when $D\le 2$. In $(2+1)$ dimensions it is possible to
write the
Riemann tensor completely in terms of the Einstein tensor
\be
R_{\mu\nu\rho\tau} \equiv
\epsilon_{\mu\nu\gamma}\epsilon_{\rho\tau\sigma}G^{\gamma\sigma}
\label{2.2}
\ee
and so a vanishing of the latter necessarily implies a vanishing of the former.
In $(1+1)$
dimensions the situation is even more extreme: the Einstein tensor vanishes for
all metrics,
and so adoption of the Einstein equations as the foundation of gravitational
theory
in $(1+1)$ dimensions necessarily implies the vanishing of the stress-energy
tensor!
One might therefore
superficially conclude that there can be no interesting gravitational
physics in lower dimensions. Fortunately the actual situation is considerably
more
interesting.

An early study of $(2+1)$ dimensional gravity \cite{DesJack} revealed
that although spacetime was
flat outside of matter, matter sources exert a gravitational influence that is
topological
in character: a point source will introduce a conical deficit into spacetime.
Consequently
vacuum gravity is only locally flat.  There is no  Newtonian limit to the field
equations and a fluid in hydrostatic equilibrium will attain its maximal mass
\cite{Cornish}. These features arise as a consequence of the comparative
mathematical
simplicity relative to the $(3+1)$ dimensional case, and in recent years this
simplicity has
been exploited to develop a formulation of quantum gravity.  This program has
been quite
successful; so successful that in fact there are six different formulations of
quantum
gravity in $(2+1)$ dimensions whose relationship remains an ongoing subject of
research
\cite{carlip}.  One
of the more recent developments in the subject has been the realization that
black holes
can exist if a negative cosmological constant is introduced \cite{BTZ}, and
that they can
form as the endpoint of collapse of a disk of dust \cite{rosscoll}.

The triviality of the Einstein tensor in $(1+1)$ dimensions necessitates a more
subtle
approach to gravitation in this case. One such approach involves rewriting the
Newtonian constant of gravity as $G_{D+1} = (1-D)\hat{G}_{D+1}/2$, where
$\hat{G}_2$ is
finite. Upon rewriting the Einstein equations (\ref{2.1}) into their trace and
trace-free parts,
it is possible to show \cite{rossred} that as $D\to 1$, the trace-free
parts simply yield $0=0$ whereas
the trace part becomes
\be
      R = 8\pi\hat{G}_2 T     \label{2.3}
\ee
where $T = T^\mu_\mu$ is the trace of the conserved $(1+1)$-dimensional stress
energy.
The above theory generalizes an early attempt at formulating a theory
of gravity in two spacetime
dimensions \cite{JackTeit} (in which the Ricci scalar was set equal to a
constant)
and was proposed several years ago \cite{MST,MFound} as a natural
classical analogue of general
relativity. The Newtonian limit exists \cite{Arnold}, and a number of $(3+1)$
dimensional features of general relativity have analogues in the theory
described by
(\ref{2.3}) \cite{RGRG,MST,Rtstuff}.  Other approaches toward formulation
of gravity in two spacetime dimenstions have involved the consideration of
non-local actions \cite{Polya}, higher-derivative theories \cite{Higher}, or
setting the one-loop beta
functions of the bosonic (or supersymmetric) non-linear sigma model to zero in
a
two-dimensional  target space \cite{MSW}.  In each
approach the Ricci scalar becomes a non-vanishing function of the
co-ordinates over some region of spacetime, permitting the spacetime to
develop interesting features such as black-hole horizons and
singularities. Considerable progress can be made in the quantization of such
theories
coupled to matter \cite{DDK,Banks,2dqgm,MLiou}, providing an attractive arena
for testing ideas about
quantum gravity \cite{GabJack}.

The field equations of virtually all $(1+1)$-dimensional theories
of gravity can be derived from
an action of the form \cite{Banks,rtDil}
\be
S = S_G + S_M = \int d^2x\sqrt{-g}\left(
\frac{1}{8\pi \hat{G}_2}[H(\psi)\frac{1}{2} g^{\mu\nu}\grad_\mu\psi
   \grad_\nu\psi +D(\psi) R] + {\cal L}_M(\psi,\Phi)\right) \label{2.4}
\ee
where the first two terms form the gravitational part of the action and the
remainder is the matter
action.  The scalar field $\psi$ is called the dilaton, and $\Phi$ denotes the
presence
of all other matter fields. The quantity
\be
{\cal J}_\mu \equiv \frac{1}{\sqrt{-g}}\frac{\delta S}{\delta
g^{\mu\nu}}\epsilon^{\nu\tau}
\nabla_\tau{\cal F} \label{2.5}
\ee
where
\be
{\cal F} =  F_0 \int^\psi ds D^\prime e^{-\int^s dt \frac{H(t)}{D^\prime(t)}}
 \label{2.6}
\ee
is conserved  ($\nabla^\mu{\cal J}_\mu = 0$) regardless of
whether or not the field
equations are satisfied, provided ${\cal L}_M$ has no metric
dependence \cite{rtDil} ($F_0$ is a constant whose value is set by the
boundary conditions on the spacetime). If a timelike
Killing vector $\xi^\mu$ exists, then one can define a conserved mass
current density \cite{rtDil}:
\be
{\cal M} = \frac{1}{2}\left[(\grad D)^2 e^{-\int^s dt
\frac{H(t)}{D^\prime(t)}}
- F_0 \int dD V e^{-\int^s dt \frac{H(t)}{D^\prime(t)}}\right]
\quad  \label{2.7}
\ee
whose space integral yields the total mass associated with
a given (stationary) solution to the
field equations.

By reparametrizing the metric and dilaton fields it is possible when
${\cal L}_M = 0$ to
convert any given $H(\psi)$ and $D(\psi)$ to any other $H(\psi)$
and $D(\psi)$.  Hence it is the
matter action which introduces interesting physics, and a given choice of
$H(\psi)$ and $D(\psi)$
describes which metric couples to matter in two spacetime dimensions. I shall
employ
several examples to illustrate this point. In all of these the metric is
written in the
coordinates:
\be
ds^2 = -\alpha(x)dt^2 + dx^2/\alpha(x) \label{2.7a}
\ee
where $\alpha$ shall be referred to as the metric function.

\subsection*{$R=T$ Theory}

This case corresponds to the choice $H(\psi)=\frac{1}{2}$,
$D(\psi)=\psi$,  and
${\cal L}_M={\cal L}_M(\Phi)$,
In this case the matter action is independent of the dilaton. This
choice for $H(\psi)$ and
$D(\psi)$ uniquely yields (\ref{2.3})
(with $8\pi\hat{G}_2 \equiv \kappa$), where the stress-energy tensor
\be
T_{\mu\nu} \equiv  \frac{1}{\sqrt{-g}}\frac{\delta S_M}{\delta g^{\mu\nu}}
\label{2.8}
\ee
is conserved. There is also an auxiliary equation for the dilaton
\cite{rtDil}.
The full system
of equations is such that the evolution of the gravity/matter system is
(classically)
unaffected by the evolution of the dilaton, although the converse is not true.

Except for being independent of the dilaton, the choice of matter action is as
arbitrary as in $(3+1)$ dimensions, being constrained by conservation laws
and positivity of energy.
One can choose it to be a two-dimensional analogue of any desired corresponding
$(3+1)$
dimensional quantity; a number of implications of this have been explored in
the
literature {\cite{MST,Rtstuff}. An interesting choice for black hole physics is
\cite{MLiou}
\be
S_M = \int d^2x\sqrt{-g} [ b(\grad\phi)^2 + \Lambda e^{-2a\phi}- \gamma \phi R]
\label{2.9}
\ee
which is the action for a Liouville field in curved spacetime. The field
equations have
the exact solution
\begin{eqnarray}
\alpha(x) &=& 1-{\lambda^2\over \mu^2}e^{-2mx} \nonumber\\
\phi(x) &=&  {m\over a}x-{1\over a}\log{m\over \mu}  \label{2.9a}\\
\psi(x) &=&   2mx-2\log{m\over \mu}+\kappa
d\log{m_0\over\lambda} \nonumber
\end{eqnarray}%
where $b = -\frac{2a^2}{\kappa}$, $d=\gamma/a-2/\kappa$,
$\lambda^2=-\Lambda/2d$ and $\mu$ and $m_0$ are positive constants.
The signs of $d$ and $\Lambda$ must be chosen
so that $\lambda$ is real. Note that the choice of constants and
origin of coordinates differs from that of ref. \cite{MLiou}.
More general choices of the coupling constants $a$ and $b$ yield other exact
solutions
to the field equations \cite{MLiou}.

\subsection*{2D String Theory}

This case corresponds to the choice $H(\psi)=4\exp[-2\psi] = 4 D(\psi)$, and
${\cal L}_M= Q^2\exp[-2\psi]$, and corresponds to the effective action of low
energy
bosonic string theory in two spacetime dimensions; the coupling
constant $Q$ may be given in terms of the central charge of the theory
\cite{MSW}.   The field equations have the exact
solution
\begin{eqnarray}
\alpha(x) &=& 1- \frac{2m}{Q} e^{-Q(x-x_0)} \nonumber\\
\psi(x) &=&    -\frac{Q}{2}(x-x_0)  \label{2.9b}
\end{eqnarray}
in the absence of any additional matter fields. These can be included
in the theory if desired.

\subsection*{Generalized String-Inspired Theories}

A natural general\-ization of the previous case in\-volves tak\-ing
$H(\psi)= 4\exp[-2\psi]$ $= 4D(\psi)$, and
  \be
  S_M = \int\sqrt{{\mbox - \,} g}\left[\,{\rm e}^{{\mbox - \,} 2\,\psi}\,
   \left( - \frac{1}{4}\,F_{\beta \sigma}\,F^{\beta \sigma} + Q^2\right)
	+ { \hspace{1mm} } \sum^{k}_{n=2} a_n\,{\rm e}^{2\,(n-1)\,\psi}
    - 8\,\pi\,{\cal L}_M\,\right] { \hspace{2mm}, } \label{2.9c}
  \ee
where the $\{ a_n\}$ are dimensional coupling constants and $F_{\mu\nu}$ is the
electromagnetic field strength \cite{Nappi}. The field equations have the exact
solution
\begin{eqnarray}
    f & = & F_{t x}  = q\,{\rm e}^{2\,\psi} { \hspace{2mm}, } \label{S2E3A} \\
    \psi & = & {\mbox - \,} \frac{Q}{2}\,(x - x_o) { \hspace{2mm}, }
\label{S2E3B} \\
    \alpha(x) & = &
    \frac{2}{Q}\,\left[\,{\cal M}(x) - m\,\right]\,{\rm e}^{2\,\psi}
    { \hspace{2mm}, } \label{S2E3C} \\
    {\cal M}(x) & = &
    \frac{Q}{2}\,{\rm e}^{{\mbox - \,} 2\,\psi(x)}\,
    \left[\,1 + \frac{q^2}{2\,Q^2}\,{\rm e}^{4\,\psi(x)}
    - \frac{1}{Q^2}\,\sum^{k}_{n=2} \frac{a_n}{n-1}\,
    {\rm e}^{2\,n\,\psi(x)}\,\right] { \hspace{2mm}. } \label{S2E3D}
  \end{eqnarray}

The spacetime may be said to have a black hole if there exists a region where
$\alpha<0$.
In the coordinates of (\ref{2.7a}) the zeros of $\alpha$ yield
the locations of the event horizons;
regions of positive $\alpha$ are regions where $\partial/\partial t$
is a timelike Killing vector.

Many other examples of two dimensional black holes exist whose
properties have been explored
to to varying degrees in the literature \cite{Lemos,Branden,Teo,Kchan}.

I would be remiss to close this section without remarking on the chief
limitations of
lower-dimensional theories of gravity: namely the absence of
tidal forces and gravitational waves in the vacuum. Since $(1+1)$
dimensional gravity is typically formulated in terms of a dilaton
field, the former limitation is perhaps not so serious, since all
interesting solutions have a non-vanishing dilaton field. However
helicity-2 gravitational waves propagating in regions of
spacetime that are free of stress
energy is one of the key features of general relativity, and is the
last significant
prediction of Einstein's theory which remains to be
 explicity verified (the binary pulsar PSR 1913+16 providing indirect
verification of this phenomenon). The absence of this feature
in lower dimensional theories
of gravity has caused a number of authors to question their
relevance for $(3+1)$ dimensional
physics. However the last decade of research has time and
again highlighted many features of
lower dimensional gravitation and spacetime structure
that are expected to survive in some
form in the real $(3+1)$ dimensional world.  Combined with the significant
payoff  in computational progress
that results from the mathematical simplicity of such theories,
this seems to me to
more than warrant as thorough an investigation of this subject as is possible.
Of course, as with any
toy model of the real world, a healthy dose of caution is advised in
extrapolating results.

\section{Outside Looking In}

In this section I shall review the main exterior properties
of lower-dimensional black holes. These
include their formation from gravitational
 collapse, their mass, energy and angular momentum,
their thermodynamic properties and their quantum properties.

\subsection*{Gravitational Collapse}

It is a well-known phenomenon in $(3+1)$ dimensions that under certain
circumstances
gravitational forces can overwhelm all other forces,
causing complete gravitational collapse
of a given system of matter \cite{HawkEll}, the end result of which is a black
hole.
The astrophysical implications of this process continue to be an
ongoing
subject of research.
That this phenomenon can also take place in lower dimensions
indicates that lower
dimensional black holes bear more than a superficial
resemblance to their higher dimensional
cousins.

Consider first the situation in $(2+1)$ dimensions.  A
circularly symmetric metric has the
general form
\be
ds^2 = -A(r)dt^2 +B(r)dr^2 + r^2 d\theta^2 \label{3.1}
\ee
and the Einstein equations (\ref{2.1}) with $T_{\mu\nu} = \frac{\Lambda}{8\pi
G} g_{\mu\nu}$ have the form
\begin{eqnarray}
-\frac{1}{2}\frac{A^\prime}{rA} - \Lambda B &=& 0 \nonumber\\
-\frac{1}{2}\frac{A B^\prime}{rB^2} + \Lambda A &=& 0 \label{3.3}\\
-\frac{1}{2}r^2{A^{\prime\prime}} - \Lambda r^2 &=& 0 \nonumber
\end{eqnarray}
where the prime denotes a derivative with respect to $r$ and $\Lambda$ is the
cosmological
constant.  The first two of these equations
yield $A(r)=1/B(r) = C - \Lambda r^2$, where $C$ is a constant of integration.
The remaining
equation yields no further constraints on $C$ (as must be the case, due to the
Bianchi identities).
Hence the Einstein equations yield the exact solution
\be
ds^2 = -(C-\Lambda r^2) dt^2 + \frac{dr^2}{C-\Lambda r^2} + r^2 d\theta^2
\label{3.4}
\ee
which is the canonical form for a deSitter metric.  Indeed, if
$\Lambda>0$, then (\ref{3.4}) is
the metric for deSitter spacetime, and $C$ must be positive for
$\partial/\partial t$ to be a timelike
direction. If $C\neq 1$ then the metric has a conical singularity at the
origin, indicative of the
presence of a mass; in the $\Lambda\to 0$ limit, we can identify $C=1-2GM$
where $M$ is
the mass associated with the conical singularity \cite{DesJack}.

If $\Lambda <0$ then the metric (\ref{3.4}) is that of anti-de Sitter
spacetime. However here a
novel feature emerges, since the sign of $C$ can be either positive or
negative. In the former case
one simply has anti-de Sitter spacetime (with a conical singularity if
$C\neq 1$). In the latter
case the spacetime has an event horizon, signalling the presence of a black
hole.
The $\Lambda >0$, $C<0$ case cannot occur for spherically symmetric metrics in
more than two spatial dimensions since the ``angular" part of the
Einstein equations force $C=1$
and the constant of integration is now $m/r^{D-2}$ where $D$ is the
number of spatial
dimensions -- the metric is that of  Schwarzchild anti de Sitter.

Given the simplicity of the above derivation it is somewhat
remarkable that the existence
of the $(2+1)$ dimensional black hole solution was uncovered by Banados {\it
et.al.}
\cite{BTZ} nearly ten years after a significant research effort in
$(2+1)$ dimensional gravity began \cite{DesJack}.
Generalizing (\ref{3.4}) to include spin, it may be shown that the metric
\be
ds^2=-N^2(r)\,dt^2+f^{-2}(r)\,dr^2+r^2\bigl(V^\phi(r)\,dt+    d\phi\bigr)^2
\label{3.4a}
\ee
where
$$N^2(r)=f^2(r)=-m+\Bigl({r\over \ell}\Bigr)^2+\Bigl({j\over2r}\Bigr)^2
   \qquad{\rm and}\qquad V^\phi(r)=-{j\over 2r^2}
$$
is equivalent to anti-de Sitter spacetime with appropriate identifications
\cite{BHTZ}.
For convenience I have set $\Lambda = -1/\ell^2$.
As with the Kerr solution, the lapse function~$N(r)$ vanishes for two
values of~$r$, namely $r_+$ and~$r_-$, where
\be
(r_\pm)^2 = {m\ell^2\over2} \pm {\ell\over2}\sqrt{m^2\ell^2 - j^2}
     \label{3.4b}
\ee
The larger of these,~$r_+$, is specified as the black hole horizon and
exists only for $m>0$ and~$|j|\le m\ell$; when~$|j|=m\ell$, $r_+=r_-$.

Recent research has indicated that there there are a wide class of
$(2+1)$ dimensional black holes \cite{Ch3dil}. These arise as exact
solutions to Einstein-Maxwell dilaton theory in $(2+1)$ dimensions. It
is also possible to show that the black hole solution (\ref{3.4a}) is
an exact solution to the Einstein equations with a topological matter
source \cite{cargeg} -- in this case $\ell$ becomes a constant of
integration whose appearance is contingent on the presence of the
topological matter fields.

Turning now to gravitational collapse,
consider a disk of collapsing dust surrounded by a vacuum region, with an
exterior metric of the form
\be
ds^2 = -(R^2/\ell^2 -M) dT^2+ \frac{dR^2}{R^2/\ell^2 - M}+ R^2d\theta^2 .
\label{3.5}
\ee
and an interior spacetime metric
\be
ds^2=-dt^2+ a^2(t) \left( \frac{dr^2}{1-kr^2}+r^2d\theta^2 \right)
\label{3.6}
\ee
which describes the freely falling dust. In these coordinates
$T_{\mu\nu} = \rho u_\mu u_\nu$ is the stress-energy of the dust, where
$\rho(t)$ is the density of the dust and $u_\mu =(1,0,0)$. Conservation of
stress-energy $T^{\mu\nu}_{\ ;\nu}=0$ then implies $\rho a^2 = \rho_0 a_0^2$,
where $\rho_0$ is the initial density of the dust and $a_0$ is the initial
scale factor. The field equations then have the solution
\be
a(t) = a_0 \cos(t/\ell) + \ell \ad_0 \sin(t/\ell), \label{3.7}
\ee
where
\be
\ad_0 = \sqrt{8\pi G\rho_0 a_0^2 - k - a_0^2/\ell^2}     \label{3.8}
\ee
yielding
\be
8\pi G\rho_0 a_0^2 - k - a_0^2/\ell^2 \geq 0.    \label{3.9}
\ee
since $a(t)$ is real. This solution always collapses to $a(t_c)=0$ in
finite proper time; when $\ad_0=0$, this is $t_c= \pi l/2$.

Imposing appropriate matching conditions to make the dust edge a
boundary surface yields \cite{rosscoll} :
\be
M = (a^2/l^2 + k +\ad^2)r_0^2-1=8 \pi G\rho_0 a_0^2r_0^2-1,  \label{3.10}
\ee
where the
dust edge is taken to be at $r=r_0$ in the interior
coordinates, and at $R= r_0a(t)$ in the exterior coordinates.
Collapse to a black hole occurs only for $\rho_0$ sufficiently large
as $M$ must be positive in this case; if $\rho_0$ is less than this
critical value, the dust collapses to a conical singularity in anti de
Sitter space.  In the case of collapse to a black hole, the curvature diverges
at
$t=t_c$, and an event horizon (which is also a surface of infinite
redshift) forms in a finite amount of exterior coordinate time
\cite{rosscoll}.

If $\Lambda \geq 0$ then collapse to a black hole is not possible. If
$\Lambda=0$ then $a(t) = a_0 + \ad_0 t$, and collapse to a conical
singularity occurs only if $\ad_0 < 0$ \cite{gidd}. If $\Lambda>0$,
then $a(t) = a_0 \cosh(\sqrt{\Lambda}t) + \frac{\ad_0}{\sqrt{\Lambda}}
\sinh(\sqrt{\Lambda}t)$ and collapse to a conical singularity is
possible only if $\ad_0 < - a_0\sqrt{\Lambda}$ \cite{rosscoll}.

These features are analogous to the $(1+1)$-dimensional case
\cite{Arnold,ross2dcoll}: a line of dust will collapse to a black hole only
if the initial density is sufficiently large.  The exterior spacetime
may be taken to be that of two Rindler spacetimes with opposite
acceleration \cite{MST,BHT} or may be taken to be the full extension
of Minkowski spacetime on either side of the line of dust
\cite{Kriele}. In either case, additional boundary conditions need to
be imposed once the endpoint of collapse has been reached.

Another form of collapse in $(1+1)$ dimensions has been studied in
the context of the string metric (\ref{2.9b}) \cite{CGHS}. In this case
leftward-moving massless scalars divide spacetime into regions: a flat
region (referred to as the linear dilaton vacuum) and a black hole
region whose metric is that of (\ref{2.9b}), with the pulse of massless
scalars at the boundary. The implications of this scenario for black
hole radiation have been extensively studied in the last two years
\cite{Strom}.

\subsection*{Energy, Mass and Angular Momentum}

The definition of mass-energy in gravitational theory is quite subtle, since
spacetime curvature
itself has stress-energy, making the localization of energy quite difficult.
An early attempt to
address this issue was developed for asymptotically flat spacetimes by
Arnowitt, Deser
and Misner \cite{ADM}. However asymptotic flatness is not always an appropriate
theoretical
idealization and is never satisfied in reality. Futhermore, although the
thermodynamic properties of black holes are expected to hold quite generally,
much of the literature on black hole
thermodynamics is restricted to the case of spacetimes that are
asymptotically flat in spacelike directions.

In recent years there has been an effort by York and collaborators to develop a
formalism
for defining
mass-energy that is independent of the assumptions of the asymptotic properties
of spacetime
curvature \cite{York,Zas}.  This approach has been referred to as the
quasilocal formalism \cite{BYork} and  can be applied to gravitational and
matter
fields within a bounded, finite spatial region, so the
asymptotic behavior of the gravitational field becomes
irrelevant (a particularly important consideration for $(2+1)$ dimensional
black holes). However even for asymptotically flat black hole spacetimes
there are certain advantages to be gained by working in a spatially
finite region: with the temperature fixed at a finite
spatial boundary, the heat capacity is positive and there is
no inconsistency in the black hole partition function\cite{Hawkgr}.  However if
the temperature is fixed at
infinity, the heat capacity for a Schwarzschild black hole is
negative  and the formal expression for the partition
function is not logically consistent \cite{Wald}.

I shall briefly review here the quasilocal formalism, highlighting those
aspects which
are pertinent for the lower dimensional cases. Further details may be
found
in ref. \cite{BCM}.

Consider a spacetime manifold~${\cal M} = \Sigma\times I$ of dimension~$(D+1)$
where $\Sigma$ is a spacelike hypersurface whose boundary is
$\partial\Sigma = B$. The boundary of~$\cal M$, $\partial\cal M$,
consists of initial and final spacelike hypersurfaces $t'$ and~$t''$,
respectively (with induced metric denoted by ~$h_{ij}$), and a timelike
hypersurface~$\Bt = B\times I$ joining these (with induced metric
{}~$\gamma_{ij}$).

The gravitational action appropriate for fixation of the
metric on $\partial\cal M$ is
\be
S^1 = {1\over\kappa}\int_{\cal M}\dDpx\sqrt{-g}(R-2\Lambda)
  +{2\over\kappa}\int_{t'}^{t''}\dDx\sqrt{h}\,K
  -{2\over\kappa}\int_\Bt \dDx \sqrt{-\gamma}\,\Theta
  \quad + S_M \label{4.1}
\ee
where $\kappa = 8\pi G$ and $\int_{t'}^{t''} d^{D}x$ denotes the difference of
integrals over the boundary elements $t''$ and $t'$. $K$ is the trace of the
extrinsic
curvature~$K_{ij}$ for $t'$ and~$t''$, defined
with respect to the future pointing unit normal and $\Theta$~is
the trace of the extrinsic curvature~$\Theta_{ij}$ of the boundary
element~$\Bt$, defined with respect to the outward pointing unit normal.

Variation of (\ref{4.1})  yields
\begin{eqnarray}
\delta S^1&=&\hbox{(terms that vanish when the equations of
motion hold)} \nonumber\\
&\qquad& +\int_{t'}^{t''} \dDx\,P^{ij}\delta h_{ij} +
\int_\Bt \dDx\,\pi^{ij}\delta \gamma_{ij}
\label{4.2}
\end{eqnarray}
where by definition
\be
P^{ij} \equiv {1\over \kappa} \sqrt{h} \bigl( K h^{ij} - K^{ij} \bigr)
\label{4.3}
\ee
is the gravitational momentum and
\be
\pi^{ij} \equiv -{1\over \kappa} \sqrt{-\gamma} \bigl( \Theta \gamma^{ij}
- \Theta^{ij} \bigr) .\label{4.4}
\ee
Terms in  $\delta S^1$ that correspond to integrals over the corners
$t''\cap\Bt$  and
$t'\cap\Bt$ will not be needed in the sequel.
The functional~$S=S^1-S^0$, where $S^0$~is a
(background) functional of the metric on~$\partial{\cal M}$, yields the
classical equations of motion when the metric is fixed
on~$\partial{\cal M}$, since in that case $\delta S^0$ vanishes.
Taking for simplicity, $S^0 = S^0[\gamma_{ij}]$ only entails the replacement
of $\pi^{ij}$ by~$\pi^{ij} - (\delta S^0/\delta\gamma_{ij})$.

The next step is to foliate the boundary element~$\Bt$ into
($D-1$)--dimensional
hypersurfaces~$B$ with induced ($D-1$)--metrics~$\sigma_{ab}$.(These
boundary elements will be points if $D=1$ -- this case will be
discussed later). The  ($D$)--metric~$\gamma_{ij}$ can be written as
\be
\gamma_{ij}\,dx^idx^j=-N^2dt^2+\sigma_{ab}
(dx^a+V^a dt)(dx^b+V^b dt) \label{4.5}
\ee
where $N$ is the lapse function and $V^a$~is the shift vector.
This yields
\be
\delta S\bigr|_{\Bt}
    = \int_\Bt \dDx \sqrt{\sigma}\Bigl( -\varepsilon\delta N +
    j_a\delta V^a + (N/2)s^{ab}\delta\sigma_{ab} \Bigr) \label{4.6}
\ee
for the contribution to the variation of~$S$ from the boundary element~$\Bt$.
Here
\begin{eqnarray}
\varepsilon &=& {2\over N\sqrt{\sigma}} u_i\pi^{ij} u_j +
    {1\over\sqrt{\sigma}} {\delta S^0\over\delta N} = {2\over\kappa} k -
\varepsilon_{0}
\nonumber\\
    j_a &=& -{2\over N\sqrt{\sigma}} \sigma_{ai} \pi^{ij} u_j -
    {1\over\sqrt{\sigma}} {\delta S^0\over\delta V^a} \quad\mbox{and}\quad
j_i = {-2\over\sqrt{h}}\sigma_{ij} P^{jk} n_k - (j_{0})_i
\label{4.7}\\
  s^{ab} &=& {2\over N\sqrt{\sigma}} \sigma_{i}^a \pi^{ij} \sigma_j^b
    - {2\over N\sqrt{\sigma}} {\delta S^0\over\delta\sigma_{ab}}
= {2\over\kappa}\bigl( k^{ab} + (n_\mu a^\mu - k)\sigma^{ab}
    \bigr)  - (s_{0})^{ab}  \nonumber
\end{eqnarray}
where $k_{ab}$~is the extrinsic curvature of~$B$ considered
as the boundary~$B=\partial\Sigma$ of a spacelike hypersurface~$\Sigma$
whose unit normal~$\boldmath{ u}$ is orthogonal to~$\boldmath{ n}$,
$P^{ij}$ denotes the gravitational momentum for the hypersurfaces~$\Sigma$
that are `orthogonal' to~$\Bt$, and  $a_\mu=
u^\nu\nabla_\nu u_\mu$ denotes the acceleration of the unit normal~$u_\mu$
for this family of hypersurfaces. Terms
proportional to the functional derivatives of~$S^0$ are respectively
denoted as $\varepsilon_0$,
$(j_0)_i$, and~$(s_0)^{ab}$.

{}From its definition $-\sqrt{\sigma}\varepsilon$ is
equal to the time rate of change of the action, where changes in time are
controlled by the lapse function~$N$ on~$\Bt$. Thus,
$\varepsilon$~is identified as an energy surface density for the system
and the total quasilocal energy is defined by integration over a
($D-1$)--surface~$B$:
\be
E = \int_B \dDmx \sqrt{\sigma}\varepsilon  \label{4.8}
\ee
Similarly, $j_i$ is the momentum surface density and $s^{ab}$ is
the spatial stress.  When there is a Killing vector
field~$\boldmath{\xi}$ on the boundary~$\Bt$,
then
\be
Q_\xi=\int_{B } \dDmx \sqrt{\sigma} (\varepsilon u^i+j^i) \xi_i
   \label{4.10}
\ee
is conserved in the sense that it is
independent of the particular surface~$B$ (within~$\Bt$) that is
chosen for its evalutation (provided there is no matter stress--energy in the
neighbourhood
of~$\Bt$). This property is not shared by the  energy~$E$.

If the system contains a rotational symmetry given by a
Killing vector field $\xi^i = \zeta^i = (\partial/\partial\phi)^i$ then
\be
J = Q_\zeta = \int_B \dDmx \sqrt{\sigma} j_i \zeta^i \label{4.11}
\ee
where the ($D-1$)--surface~$B$
is chosen to contain the orbits of~$\boldmath{\zeta}$.
If the Killing vector field~$\boldmath{\xi}$ is timelike, then the negative of
the corresponding charge~(2.18) defines a conserved mass for the system,
\be
M = - Q_\xi = \int_B \dDmx \sqrt{\sigma} N\varepsilon \label{4.12}
\ee
the latter equality holding if the
Killing vector field is also surface forming, where $B$ has a unit normal
proportional to~$\boldmath{\xi}$ and
$N$ is the lapse function defined by~$\mbox{$\boldmath{\xi}$} =
N \mbox{$\boldmath{u}$}$.
If~$\boldmath{\xi}$ does not have unit norm at~$B$, then the mass~$M$
will differ from the energy~$E$. In general the energy~$E$ evaluated
on a given slice of~$\Bt$ will
not equal the conserved mass~$M$.

These distinctions  between mass and energy are especially
important for spacetimes that are
not asymptotically flat. In the  anti-de Sitter  case the magnitude of
the timelike Killing vector field diverges as it approaches infinity.
It does not approach
the unit normal to the (asymptotically) stationary time slices at spatial
infinity,
and the mass~$M$ and energy~$E$ do not coincide. Explicitly the
metric in $(3+1)$ dimensions
is
\be
ds^2 = -N^2(r) \,dt^2 + f^{-2}(r) dr^2     +
r^2(d\theta^2+\sin^2\!\theta\,d\phi^2)  \label{4.13}
\ee
where $ N^2(r) = f^{2}(r) = 1-2m/r + r^2/\ell^2$. Choosing the
boundary $B$ to be a surface of
constant $r=R$, the extrinsic curvature is $k = -{2f(R)\over R}$
and (\ref{4.8}) and (\ref{4.12})
respectively give
\be
E = -\frac{16\pi}{\kappa}R\sqrt{1 - 2m/R + R^2/\ell^2} - 4\pi
R^2\varepsilon_0(R)  \label{4.14}
\ee
\be
M = N(R) E \label{4.15}
\ee
Choosing $\varepsilon_0(R)  = -{4\over\kappa R} \sqrt{1 + R^2/\ell^2}$
({\it i.e.} the reference spacetime is anti-de Sitter space, so that $E$
vanishes when $m=0$)
yields as $R\to\infty$ a vanishing energy $E\sim m\ell/R \to 0$ but finite mass
$M\to m$.

Turning now to the $(2+1)$ dimensional black hole, the metric is given by
(\ref{3.4a}).
Specifying $B$ to be a surface of constant $r=R$, it is straightforward to show
\cite{BCM}
that $k = -{f(R)\over R}$ and that (\ref{4.8}) and (\ref{4.12}) now yield
\be
E = -\frac{4\pi}{\kappa} \sqrt{-m + R^2/\ell^2 + j^2/(4R^2)} -     2\pi
R\varepsilon_0(R) \label{4.16}
\ee
\be
M = N(R)\,E + {J^2\over 2R^2} \label{4.17}
\ee
and that
\be
J = j  \label{4.18}
\ee
from (\ref{4.17}). The energy~$E$ and angular momentum~$J$ will
vanish for the zero mass
black hole (the metric (\ref{3.4a}) with~$m=0$, $j=0$) if  the choice
$\varepsilon_0(R) = -{2\over \kappa\ell} $ is made; as~$R\to\infty$ this
implies
$E\sim m\ell/R \to 0$ and $M\to m$.

These results are consistent with
an analysis of the Hamiltonian for (2+1) gravity which shows that $m$ and~$j$
are the ADM mass and angular momentum at infinity \cite{BHTZ}.
They are also consistent with the formulation of $(2+1)$ gravity
as a Chern-Simons gauge theory of the Poincare group
\cite{carlip}; in this case the parameters $m$ and
$j$ may be interpreted in terms of Casimir invariants of mass and spin
respectively \cite{CLM}.

The methods described here can also be applied straightforwardly to
the $(2+1)$ dimensional black hole solutions discussed in refs.
\cite{Ch3dil,cargeg}. For the dilaton solutions one finds that there
exists a well defined quasilocal mass at large $R$ that is
proportional
to the expected constant of integration in the field equations \cite{Ch3dil}.
However for the topological $(2+1)$ black hole \cite{cargeg} one finds
the rather surprising result that the mass is proportional to the
parameter $j$ and the angular momentum is proportional to the
parameter $m$! The implications of this are still under investigation.

In $(1+1)$ dimensions the development of the quasilocal formalism \cite{2djol}
begins with the action
(\ref{2.4}), modified to include boundary terms of the form
\be
S_{\rm B} =
\int\bigl[-2\epsilon\sqrt{-g}\nabla_a\bigl(D(\psi)Kn^a\bigr)\bigr]\label{4.19}
\ee
where $n^a$ is a unit vector, $n^an_a=\epsilon$ with $\epsilon=\pm1$
for spacelike/timelike $n^a$. The extrinsic curvature $K_{ab}$ is defined
as $K_{ab}=-h^c_a\nabla_cn_b$, where
$h_{ab}=g_{ab}-\epsilon n_an_b$ and  $K=g^{ab}K_{ab}=h^{ab}K_{ab}$.
An analysis similar to the one described above can be performed, the
chief difference being that the $(D-1)$ dimensional boundary $B$ now
consists of either one or two points (depending on the choice of
boundary). Although there is no formula analogous to (\ref{4.11})
(since
there is no angular momentum in $(1+1)$ dimensions), the
remaining formulae in the $D=1$ case are
\be
E =-\frac{2}{\kappa}\bigl(n^a\nabla_aD(\psi)\bigr)-{\cal E}_0  \label{4.20}
\ee
and
\be
M =\frac{2}{\kappa}(u\cdot\xi)\bigl(n^a\nabla_aD(\psi)\bigr)- M_0  \label{4.21}
\ee
for the quasilocal energy and mass respectively, where $\boldmath{\xi}$ is a
timelike
Killing vector. Applying these results to the $(1+1)$ dimensional metrics
discussed in
the previous section yields for large $x$
\be
E=M \to 2dm \label{4.22}
\ee
for the metric (\ref{2.9a}), and
\be
E=M \to 2m/\kappa \label{4.23}
\ee
for the metric (\ref{2.9b}). Note that if $\gamma\neq 0$ in (\ref{2.9}), the
formula
in (\ref{4.21}) must be modified so that $D(\psi)\to D(\psi,\phi)$;
provided no additional surface terms are added to the
action, equation (\ref{4.23}) yields the same result as (\ref{2.7})
\cite{MLiou,2djol}.
Note that $E=M$ at large $x$ since these spacetimes are
asymptotically flat.

\subsection*{Thermodynamic Properties}

The methods of the previous section may now be employed to analyze the
thermodynamic
behaviour of lower dimensional black holes.

Consider first the temperature. For a spherically symmetric metric,
it is given by \cite{York,BCM}
\be
T = \frac{1}{2\pi N(R)} {\kappa_H}\label{5.1}
\ee
where $R$ is the boundary radius as described in the previous section and where
\be
\kappa_H^2 = \frac{1}{2} (\grad^\mu\xi^\nu)(\grad_\mu\xi_\nu) \vert_H
\label{5.2}
\ee
is the surface gravity at the event horizon; $\xi^\mu$ is a timelike Killing
vector outside
the horizon.
The factor $N(R)$ is the redshift factor. It will approach unity in
asymptotically flat spacetimes, but will diverge in anti-de Sitter spacetime.

For the $(2+1)$ dimensional black hole the temperature is \cite{BCM}
\be
T = {1\over \sqrt{-m + R^2/\ell^2 + (J/2R)^2} }\biggl(
    {(r_+/\ell)^2 - (J/2r_+)^2 \over r_+} \biggr) \label{5.3}
\ee
where this result may also be derived by taking $T \equiv {\partial
E\over\partial{\cal S}}$, where
$\cal S$ is the entropy of the black hole.

The entropy may be calculated using either the first law of thermodynamics or
by computing the
surface integral of the Noether charge associated with diffeomorphism
invariance \cite{Wald}. This latter method may be used to show that the entropy
of the
(2+1) black hole~(4.1) is~$4\pi r_+/\kappa$: twice the `area' of its event
horizon \cite{BTZ}. These results may further be employed to evaluate the heat
capacity of the
black hole \cite{BCM}.   At constant surface `area'~$2\pi R$ and constant
angular momentum~$J$  this quantity is
\be
C_{R,J} \equiv \biggl( {\partial E\over\partial T}\biggr) =
    \biggl( {\partial E\over\partial r_+} \biggr)
    \biggl( {\partial T\over\partial r_+} \biggr)^{-1} \label{5.4}
\ee
where the energy and temperature are expressed as functions
of $r_+$, $R$, and~$J$. It is straightforward to show that
$\partial T/\partial r_+$ is positive, so that the temperature
is a monotonically increasing function of~$r_+$. This means that
there is a unique black hole with a given temperature~$T(R)$
and a given angular momentum~$J$.

In $(1+1)$ dimensions for metrics of the form (\ref{2.7a}) the temperature is
given by
\cite{MST,2djol}
\be
T = \frac{1}{4\pi}{\alpha^\prime(x_H)}{\cal R}(x)\label{5.5}
\ee
and the entropy is
\be
{\cal S} = \frac{4\pi}{\kappa} D\bigl(\psi(x_{\rm H}) \bigr) \label{5.6}
\ee
where ${\cal R}(x)$ is a redshift factor whose value will depend upon the lapse
function. In the large $x$ limit these quantities are respectively
\be
T = \frac{M}{4\pi d} \qquad  {\cal S} =  \frac{4\pi}{\kappa} d\ln(M/M_0)
\label{5.7a}
\ee
for the metric (\ref{2.9a}) (where $M_0=2d m_0$) and
\be
T = \frac{Q}{4\pi} \qquad  {\cal S} = \frac{2M}{Q} \label{5.7b}
\ee
for the metric (\ref{2.9b}).

Despite the similar functional form of the metrics (\ref{2.9a}) and
(\ref{2.9b}), the thermodynamic
properties of the two metrics are markedly different. For the Liouville black
hole (\ref{2.9a}) the temperature is proportional to the mass and the entropy
varies logarithmically with the mass,
whereas in the string-inspired case (\ref{2.9b}) the temperature is constant
and the entropy varies linearly with the mass.

I shall close this section by considering some of the implications of a black
hole whose
entropy varies logarithmically with the mass \cite{TomRobb}. Some insight into
the
properties of such a black hole can be gained by considering its behaviour
inside a box of radiation, a non-trivial problem in $(3+1)$ dimensions
\cite{PCW,GP}.
In this case a black hole contained in a perfectly
reflecting cavity filled with thermal energy will not be able to achieve
thermal
equilibrium if the cavity is sufficiently large.

For the analogous $(1+1)$ dimensional system, the metric is taken to be that
of (\ref{2.9a}), with $x \to |x|$; such a metric would describe the exterior of
a
collapsing line of matter in the background of a Liouville field -- this metric
can be
extended everywhere that $x\neq 0$, where there is a delta-function singularity
in
the curvature scalar.  Consider a box of length
$L$ containing a black hole of mass $M$ and thermal radiation at temperature
$T$,
with $mL >> \ln(\lambda/\mu)$.
The energy and entropy of this system are given by
\begin{equation}
{\cal S}=2\pi \ln \Bigl( {M\over M_0}\Bigr) + {\pi\over 3} TL
{}~~~~~~~E=M+{\pi\over 6} T^2L  \label{5.8}
\end{equation}
where the latter term arises from the Stefan-Boltzmann law in two spacetime
dimensions
and $2d$ has been set to unity for simplicity. Maximizing the entropy for
a fixed energy $E$ (the microcanonical ensemble) yields
\begin{equation}
0={\partial {\cal S}\over \partial M}={2\pi\over M} - {1\over T} ~~
\Rightarrow ~~T={M\over 2\pi} \label{5.9}
\end{equation}
providing an alternative derivation of (\ref{5.7a}). It is straightforward to
show that
${\partial ^2{\cal S}\over \partial M^2} < 0 $, guaranteeing a maximum.

Hence as long as the equilibrium condition (\ref{5.9}) can be physically
realized
the entropy is always maximized.  Clearly this cannot be trivial -- indeed,
insertion of (\ref{5.9}) into (\ref{5.8}) implies
\begin{equation}
ML=-12\pi +12\pi\sqrt{1+{EL\over 6\pi}} >> 2\ln(m_0/\lambda)\quad .
\label{5.10}
\end{equation}
the latter inequality guaranteeing
that the horizons are contained within the box, thereby constraining
$EL$ above a certain threshold.

Comparing the relative values of the entropies
of a system of pure radiation and the combined black-hole/radiation system, it
is possible
to show that a phase transition can occur between the two systems
\cite{TomRobb}
For a black hole contained in a box of length $L$ which respects the
equilibrium condition (with $ML$ is given by (\ref{5.10}))
\begin{equation}
{\cal S}_{bh}=2\pi\ln\Bigl( {ML\over M_0L}\Bigr) + {1\over 6}ML  \label{5.11a}
\end{equation}
whereas for pure radiation contained in a box
of the same size
\begin{equation}
{\cal S}_{rad}=2\sqrt{{1\over 6}\pi EL} \quad .
\label{5.11b}
\end{equation}
Fixing the value of $M_0L$, it may be shown that the temperature
initially increases with energy and the entropy is maximized by pure
radiation.  A phase transition is then reached, forming a black hole
accompanied by a sharp drop in temperature \cite{TomRobb}.

It is furthermore possible to  show that the functional form of the entropy
\be
{\cal S}=2\pi \ln \Bigl( {M\over M_0}\Bigr) \label{5.12}
\ee
implies that it is thermodynamically favourable for lower-dimensional black
holes to
fragment \cite{TomRobb,ptl}.
Consider a $(1+1)$-dimensional system consisting of a black hole of mass
$M$ which has spawned $n$ identical black holes of mass $m$. The
entropy is
\be
{\cal S}_n = 2\pi \ln\left[\frac{M-nm}{M_0}\left(\frac{m}{M_0}\right)^n\right]
\label{5.13}
\ee
provided the subsystems are sufficently separated
to be regarded as independent. Eq. (\ref{5.13}) becomes
(\ref{5.12}) for $n=0$. It is straightforward to show that ${\cal S}_n$ is
maximized if
$m=M/(n+1)$, implying
\be
{\cal S}^{\rm max}_n = 2\pi (n+1)\ln\left[\frac{M}{(n+1)M_0}\right] \quad .
\label{5.14}
\ee
When $n=1$ the system of separated black holes has larger entropy than that of
a single
black hole provided $M/M_0 >4$; the situation is reversed if $M/M_0 >4$.
For $n$ large enough to permit differentiation of ${\cal S}^{\rm max}_n$
with respect
to $n$, one finds (given $M$ and $M_0$)
its maximal value ($\hat{\cal S}_{n_0}$, say)  to be at $n=n_0$, where
\be
n_0 = \frac{M}{e M_0} -1 \quad\mbox{and}\quad
\hat{\cal S}_{n_0} = 2\pi \frac{M}{e M_0} = 2\pi(n_0 + 1) \quad . \label{5.15}
\ee
Strictly speaking $n_0$ must be an integer and $\hat{\cal S}_{n_0}$ is a
rising staircase curve as a function of $n_0$.

To find the entropically  best transition for a black hole of mass M
one must ensure that ${\cal S}_n > {\cal S}_0$, or alternatively $M/M_0 >
(n+1)^{1+1/n}$. This yields the number of equal black holes into which the
original one will fragment.  Setting ${\cal S}_0 = {\cal S}_1$
determines the value of $M$ for which there is no gain in entropy for the
black hole system to either merge or fragment.
This occurs for
\be
\frac{M}{M_0}=4 \qquad\mbox{where}\quad {\cal S}_0={\cal S}_1 =2\ln(2)=1.386
\quad .
\ee
For no fragmentation $M/M_0< 4$ and
the fragmentation process always comes to a stop, as one
intuitively expects.

Lower dimensional black holes also give rise to several interesting points
of principle in thermodynamics which are discussed in ref. \cite{ptl}.

\subsection*{Quantum Properties}

The study of the quantum properties of lower dimensional black holes has been
the subject of
intense research for the past two years. Much of the attention has focussed on
the model
described by (\ref{2.9b}), since significant progress can be made in solving
the back-reaction
problem associated with black hole evaporation. A full discussion of this
subject is beyond
the scope of this paper and I refer the reader to a recent review of the
subject \cite{Strom}.

A recent study of Liouville black holes has shown that it is possible to make
significant progress
on the back-reaction problem in this case as well \cite{MLiou}.
These metrics are solutions to the field equations of  the $R=T$
theory, with matter action (\ref{2.9}). A wide class of such
solutions exists, one of which is the metric (\ref{2.9a}). Coupling
this system to a set of conformally invariant matter fields with
central charge $c_M$
and evaluating the path integral of
to one-loop order yields the result that the space of solutions
maps into itself, so that
a given classical solution
\be
ds^2 = -\alpha(x;G) dt^2 +  \frac{dx^2}{\alpha(x;G)}     \label{5.16}
\ee
is mapped to
\be ds^2 = -\alpha(x;G_R) dt^2 + \frac{dx^2}{\alpha(x;G_R)}
\label{5.17}
\ee
where
\be
G_R = \frac{G}{1- \frac{c_M\hbar G}{3}} \label{5.18}
\ee
is the renormalized gravitational constant. A more
detailed study of this system is presently
under investigation.

\section{Inside Looking Out}

The realization that a black hole can form from gravitationally collapsing
matter naturally
leads to the question of what the final fate of the collapsing matter is.
In $(3+1)$ dimensions it has been demonstrated that the
spacetime exterior to a collapsing body relaxes to that of a Kerr-Newman
(KN) black hole, with radiative perturbations decaying as advanced time
increases according to a power law, provided the (unproven) hypothesis of
cosmic censorship is valid.

The interior situation is much less well understood.
Infalling matter will either encounter a spacelike region of diverging
curvature (at which point quantum gravitational effects presumably
dominate) or alternatively will avoid the singularity and emerge into
another universe via a `white hole', the prototypical case being the KN
geometry.
However the stress-energy associated with
massless test fields diverges at a null hypersurface inside the black hole
called the Cauchy horizon \cite{Penrose}, suggesting an instability in the
interior
geometry. Any object falling into a KN
black hole must eventually cross the Cauchy horizon, and so an
understanding of its stability is intimately connected to the question of
the final fate of the infalling matter.

Further insight into the fate of the interior was made several years ago
by Poisson and Israel
\cite{Poisson}, who demonstrated that the Cauchy horizon of the
Reissner-Nordstr\"{o}m solution forbids any evolution of spacetime beyond
this horizon.   The mass parameter becomes unbounded inside the black hole
due to the presence of ingoing and backscattered outgoing radiation,
and the Kretsch\-mann scalar diverges. Ori \cite{Ori} later developed a
simpler model of this phenomenon, and  argued that the mass inflation
singularity was
too weak to forbid passage through the Cauchy horizon since
its tidal forces do not necessarily destroy any physical
objects. This extensibility problem remains a subject of some controversy
\cite{Bonnano}.

Most recently mass inflation has also been shown to take place in
lower-dimensional black holes \cite{JChan,Droz,Husain,CCM,Jenf}.
While several of the physical properties are quite analogous to the higher
dimensional cases
previously studied, some novel features emerge. I shall consider
separately the $(1+1)$ and $(2+1)$ dimensional cases.

\subsection*{Inside a $(1+1)$ Dimensional Black Hole}

Consider first the $(1+1)$ dimensional situation \cite{JChan},
where the action is given by
(\ref{2.9c}), with an additional term added to
the matter action for a null fluid, whose
stress-energy tensor is
  \begin{eqnarray}
    T_{\mu \nu} & = & \rho(v)\,l_{\mu}\,l_{\nu} \period \label{S2E4}
  \end{eqnarray}
In this case the field equations have the exact solution (\ref{S2E3A})
-- (\ref{S2E3D})
  \begin{eqnarray}
    \de s^2 & = & 2\,\de x\,\de v - \alpha(x,v)\,\de v^2 \comma \label{S2E2}
  \end{eqnarray}
where $\alpha(x,v)$ is given by (\ref{S2E3C}), with $m\to m(v)$.
The function $m(v)$ satisfies the differential equation
  \be
    \frac{dm}{dv} = 8\pi G \rho(v) . \label{S2E5}
  \ee
For generic non-vanishing $a_n$ the spacetime described by
the metric (\ref{S2E2})  has multiple horizons at $x_I$, where $\alpha(x_I)=0$.
The largest postive value of $x_I$ shall be called the outer horizon and the
next largest value the
Cauchy horizon.

  Consider matching two patches of solution
  (\ref{S2E3A})--(\ref{S2E3D}) along an outgoing null line.
Denoting the chronological past of the union of the null ray and the Cauchy
horizon by
region I and its complement by region II, the mass functions in the respective
regions are given by
\be
m =   m_{1}(v_{1}) = m_0  - \delta m(v_1) \quad \mbox{and} \quad  m =
m_{2}(v_{2})
\label{6.1}
\ee
where $\delta m(v_1) = h v_1^{-p}$ (in region I) models the radiative tail of
the
collapsing null matter, and has a power-law falloff in $(1+1)$ dimensions
\cite{JChan}.
The parameter $m_0$ corresponds to the final mass of the black hole.

The phenomenon of mass inflation in this context may be easily seen by
matching the metric for an influx of radiation with different mass
functions $m_1$ and $m_2$ along the outgoing
null line. These requirements imply \cite{JChan}
  \begin{equation}
    \alpha_1\,\de m_2 = \alpha_2\,\de m_1 \label{match3}
  \end{equation}
showing that near the Cauchy horizon, where $\alpha_1 \to 0$ as
  $v_1 \to \infty$, the inner mass function $m_2$ diverges, as all
  other quantities remain finite.

More explicitly, the coordinate system
  (\ref{S2E2}) implies that the outgoing null geodesic satisfies the equation
  \begin{eqnarray}
  2\,\dot{x}(\lambda) & = &
  \alpha\left(\,x(\lambda),v(\lambda)\,\right)\,\dot{v}(\lambda) \comma
  \label{S2E6}
  \end{eqnarray}
  where $\lambda$ is an affine parameter and
  the dot denotes derivative with respect to $\lambda$. Without loss of
  generality, the parameter $\lambda$ may be taken to be zero
  at the Cauchy horizon and positive beyond that.

Consider first the case of only two horizons. The matching conditions imply
that
  \begin{eqnarray}
    v_{1}(\lambda) \approx
    \minus \frac{1}{k_o}\,\int^\lambda \EXP^{2\,\psi(X_{c}) -
2\,\psi(X(\zeta))}\, \frac{\de \zeta}{\zeta}
     \approx \minus \frac{1}{k_o}\,\int^{\lambda} \frac{\de \zeta}{\zeta}
     =  \minus \frac{1}{k_o}\,\ln|\,\lambda\,| && \label{S2E16}
  \end{eqnarray}
  in some negative neighborhood of $\lambda = 0$, and
  \begin{eqnarray}
    v_{2}(\lambda) \approx
    \int^{\lambda} \EXP^{\minus 2\,\psi(X_{c})} / Z_{2}\,\de \zeta
     =  \EXP^{\minus 2\,\psi(X_{c})}\,\frac{\lambda}{Z_{2}} \period
\label{S2E17}
  \end{eqnarray}
which vanishes at the Cauchy horizon.
Here $X(\lambda)=x(\lambda)$ along the null ray
and $k_o = \minus \frac{1}{Q}\,{\cal M}'(X_c)\,\EXP^{2\,\psi(X_{c})} > 0$.
 The ($\lambda$-dependent) ``mass'' of the null particle can be
  defined as
  \begin{eqnarray}
    \Delta m(\lambda) \b := \b m_{2}(\lambda) - m_{1}(\lambda) \b = \b
    Q\, - Z_{2}\,\dot{X}(\lambda) \period \label{S2E10}
  \end{eqnarray}

Near the Cauchy horizon,  when $\lambda$ is close to zero
$\Delta m$ is approximately
\begin{eqnarray}
    \Delta m(v_{1}) & \approx &
    \frac{h\,Z_{2}}{k_o}\,\EXP^{2\,\psi(X_{c})}\,|\,v_{1}\,
    |^{\minus p}\,\EXP^{k_o\,v_{1}} \period \label{S2E20}
  \end{eqnarray}
where $Z_2 > 0$ and $m_2$ is
  \begin{eqnarray}
    m_{2}(\lambda) & = &
    M - \delta m(\lambda) + \Delta m(\lambda) \nonumber \\
    m_{2}(v_{2}) & \approx &
    M - h\,\left[\,1 + \frac{1}{k_o\,v_{2}}\,\right]\,|\,k_o\,|^p\,
    \left|\,\ln\left|\,Z_{2}\,\EXP^{2\,\psi(X_{c})}\,v_{2}\,
    \right|\,\right|^{\minus p} \comma \label{S2E21}
  \end{eqnarray}
  where (\ref{S2E16}) and (\ref{S2E17}) are used. This shows that in
  the case ${\cal M}'(X_c) \neq 0$, the mass in region II becomes
  unbounded near the Cauchy horizon where $v_2 \approx 0$.

The presence of multiple horizons can change this picture
\cite{JChan}.  Since
  our metric solutions are basically polynomials of a degree higher
  than 2, for certain values of the parameters $a_n$ in (\ref{2.9c})
it is possible for
  $k_o$ to vanish, {\it i.e.} for
  ${\cal M}'(X)$ to vanish at the (first) inner horizon. More generally, it is
  possible for the first $N$ derivatives of ${\cal M}(X)$ to be
  zero at the Cauchy horizon, where $N$ is some positive integer.

Since ${\cal M}(X)$ is a finite polynomial there will be
  some order of derivative of ${\cal M}(X)$ at $X = X_c$ that will
  be non-zero. Denoting this by $b$, such that
  ${\cal M}^{(i)}(X_c) = 0$ but ${\cal M}^{(b+1)}(X_c) \neq 0$ for
integer $b \geq 1$ such that for every integer $i \in [1, b]$,
it may be shown that (\ref{S2E16}) is modified to
  \begin{eqnarray}
    v_{1}(\lambda) & \approx & \frac{1}{b\,\hat{\kappa}}\,\lambda^{\minus b}
    \label{S4E5}
  \end{eqnarray}
where $\hat{\kappa}$ is a positive constant. Since $v_1$ no longer has
logarithmic behaviour (\ref{S2E20}) becomes
  \begin{eqnarray}
    \Delta m(\lambda) & \approx &
    \minus h\,Z_{2}\,\EXP^{2\,\psi(X_{c})}\,\hat{\kappa}^{p - 1}\,b^p\,
    \lambda^{b\,p - b - 1} \comma \label{S4E7}
  \end{eqnarray}
  where $Z_{2}$ again must be positive in order to have a positive mass.
Furthermore, as $X_c$ is the Cauchy horizon, we also have
  ${\cal M}(X_c) = M$.

  From expression (\ref{S4E7}), it is clear that two distinct possible
scenarios can occur. If $p < 1 + 1 / b$, $\Delta m$ will be
  unbounded as $\lambda \rightarrow 0$, and the inner mass parameter
  $m_{2}$ will also inflate because $m_{2} = m_{1} + \Delta m$, just
as in the previous situation.  However
if $p > 1 + 1 / b$, {\it there will not be any mass inflation at all}
  because the exponent of $\lambda$ in (\ref{S4E7}) is positive. When
  $\lambda \rightarrow 0$, the mass of the particle tends to zero when
  $p$ is large enough; the boundary particle `deflates' as it approaches
  the Cauchy horizon.

An analogue of this situation in $(3+1)$ dimensions would require the
existence of a black hole solution with vanishing surface gravity at
the inner horizon.  Whether or not such a solution can be obtained
under physically reasonable circumstances is not yet clear.

\subsection{Inside a $(2+1)$ dimensional Black Hole}

It is also possible to show that mass inflation
occurs for the $(2+1)$ dimensional black hole \cite{CCM,Jenf},
explictly including effects due to rotation.

Consider modifying the Einstein equations
(\ref{2.1}) in $(2+1)$ dimensions
by including an electromagnetic stress-energy tensor,
and the stress-energy
tensor of a rotating null fluid
 with energy density $\hat{\rho}$ and angular momentum density
$\hat{\omega}$, that is
\begin{eqnarray*}
  \left[\,{\cal T}_{\mu \nu}\,\right] & = &
  \left[\,\begin{array}{ccc}
  \hat{\rho}(v,r)          & 0 & \minus \hat{\omega}(v,r) \\
  0                        & 0 & 0 \\
  \minus \hat{\omega}(v,r) & 0 & 0
  \end{array}\,\right] \comma
\end{eqnarray*}
in addition to including a cosmological constant. The field equations
then have the exact solution \cite{CCM}
\begin{eqnarray}
  \de s^2 & = &
  \left[r^2/\ell^2 + m(v) + 4\,\pi\,G\,q^2\,\ln(r / r_o)\,\right]\,\de v^2
  \nonumber \\ && \quad
  + \s 2\,\de v\,\de r - j(v)\,\de v\,\de \theta
  + r^2\,\de \theta^2 \comma \label{E2}
\end{eqnarray}
where $m(v)$ and $j(v)$ satisfy the differential equations
$$
  \frac{\de m(v)}{\de v} = 16\,\pi\,G\,\rho(v) \quad \mbox{and} \quad
  \frac{\de j(v)}{\de v} = 16\,\pi\,G\,\omega(v)
$$
and
$\hat{\rho}(v,r) = \rho(v) / r + j(v)\,\omega(v) / (2\,r^3)$ and
$\hat{\omega}(v,r) = \omega(v) / r$, as dictated by the conservation
laws, with $\rho$ and $\omega$ arbitrary functions of $v$.

Consider a pulse, $S$, of outgoing null radiation between the Cauchy
and outer horizons in the background spacetime (\ref{E2}). We can
model this by matching two patches of solution (\ref{E2}) with
different $m$ and $j$ along $S$. The region enclosed by the
ring and its complement will be characterized by their distinct values
$m_a(v_a)$ and $j_a(v_a)$
where $a$ has value of  2 (1) for the enclosed (non-enclosed)
region. The two regions have different masses and the Cauchy
horizon cannot coincide with the inner horizon.
If $j_1 = j_2$, the null ring will rotate at the same pace as the
spacetime in both regions. However, when $j_1 \neq j_2$, $S$
will carry intrinsic spin.

As before,
continuity of inflow along a null curve with
tangent vector
$$
  l^{\s \mu} \b = \b
  \left<\,\frac{2}{N^{\s 2}} \comma 1 \comma
  \frac{j}{r^2\,N^{\s 2}}\,\right>
$$
yields
\be
  \frac{\de m_1 - \de (j_1^{\s 2}) / (4\,R^2)}{N_1^{\s 2}} =
  \frac{\de m_2 - \de (j_2^{\s 2}) / (4\,R^2)}{N_2^{\s 2}}
  \period \label{E5}
\ee
as the $(2+1)$ dimensional analogue of (\ref{match3}),
where $R(\lambda)$ is such that $2\,\pi\,R$
is the perimeter of $S$.  Taking the affine parameter $\lambda$
to be zero at the Cauchy horizon and positive behind that, and in
addition taking $M$ and $J$ to be the respective asymptotic values of
$m_1$ and $j_1$ we can again see that inflationary behaviour will
occur: when the ring is close to the Cauchy horizon $R=r_c$, $v_1$
approaches infinity and the right hand side of (\ref{E5}) diverges.

Note that the inclusion of angular momentum implies that it is
the quantity $E \equiv m(v) - j^2(v)/(4 r^2)$ which inflates.
This quantity is proportional to the total energy of spacetime
at large $r$, neglecting electromagnetic contributions \cite{BCM} as
discussed above. A more detailed analysis shows that \cite{CCM,Jenf}
\begin{eqnarray}
  v_1(\lambda) \b \approx \b \minus \frac{1}{\hat{k}_o}\,\ln|\lambda| & &
  v_2(\lambda) \b \approx \b \frac{2}{Z_2}\,\lambda \period
  \hspace{8mm} \label{E9}
\end{eqnarray}
where $\hat{k}_o$ is a positive constant and that
\begin{eqnarray*}
 E_{\rm ring}(\lambda) & \approx &
       \minus \frac{Z_2}{2\,\hat{k}_o\,\lambda}\,\delta E(\lambda) \comma
\end{eqnarray*}
where $Z_2$ must be positive so that $E_{\rm ring} > 0$. If we
assume that $\delta E(\lambda)$ decays to zero via a power law
$h\,v_1^{\s \minus p}$ \cite{Husain,CCM,Price}, we obtain
\be
  E_2(v_2) \approx
  M - \frac{J^2}{4\,r_c^{\s 2}} \nonumber
  - \s \frac{h}{v_2}\,\hat{k}_o^{\s p-1}\,
  \left|\,\ln\left|\frac{Z_2\,v_2}{2}\right|\,\right|^{\minus p}
  \period \label{E10}
\ee
As a result, $E_2(v_2)$ goes to infinity while $S$ approaches the Cauchy
horizon because $v_2$ tends to zero from below at that instant.

We close by considering tidal distortions at the horizon. In the
triad frame, the relevant components of the Riemann tensor look
like the following:
\begin{eqnarray*}
  R^1{}_{001} & = &
  \frac{\div \alpha - \dir \alpha}{2\,r} - \frac{\div (j^2)}{8\,r^3}
  - \frac{j^2}{8\,r}\,\dir \left(\,\frac{\dir \alpha}{r}\,\right) \comma \\
  R^1{}_{002} & = & R^2{}_{001} \b = \b
  \frac{j}{4}\,\dir \left(\,\frac{\dir \alpha}{r}\,\right)
  + \frac{\div j}{2\,r^2} \comma \\
  R^2{}_{002} & = & \minus \half\,\dirr \alpha
\end{eqnarray*}
where $\alpha = -g_{vv}$.
It is clear that the most divergent component is $R^1{}_{001}$.
The tidal distortion is finite since one
can approximate the distortion by integrating the above components
twice with respect to $v$ \cite{Ori,Balbinot} and obtain a finite
result.
Furthermore the Kretsch\-mann scalar of the BTZ solution is given by
$$
  R_{abcd}\,R^{abcd} =
  \frac{2}{r^2}\,\left[\,\dir \alpha(v,r)\,\right]^2
  + \left[\,\dirr \alpha(v,r)\,\right]^2
$$
which is obviously bounded at Cauchy horizon. Hence there is no reason
to  terminate the classical extension of the spacetime beyond
the Cauchy horizon.

Some of the qualitative features of this phenomenon carry over to
$(3+1)$ dimensions. It is possible to show that a black string will
also undergo mass inflation in a manner similar to that described
above \cite{BS}.

\section{Outlook}

Although research in lower dimensional black holes has mushroomed since the
advent of the subject five years ago, much remains to be done.  I shall close
by
suggesting two interesting lines of research in the subject which might be
pursued.

\begin{itemize}

\item Black Hole Radiation

To a large extent, the surge of interest in lower dimensional black holes was
caused
by the realization that such models offer the possibility of explicitly
incorporating back
reaction effects due to their mathematical simplicity \cite{CGHS,Strom}.
However to
date the physics of this problem has been studied virtually exclusively in the
context
of the metric (\ref{2.9b}).  This is a 2 dimensional black hole whose
temperature is
constant and whose entropy is proportional to the mass. As such it is a very
special
model. It would be of interest to see to what extent the physics of black hole
radiation
(including the back reaction) is dependent upon these properties. The wide
variety
of black hole metrics available in two dimensions permit such an investigation,
and this
remains largely unexplored territory.  The Liouville black hole metric
(\ref{2.9a}) provides
a most interesting contrast: as it evaporates, its temperature and entropy both
decrease, and it can reach a state of vanishing entropy at finite mass
in a finite amount of time \cite{MLiou,2djol}, reminiscent of a remnant.
Evolution of the spacetime beyond this point is at present unclear.

\item Black Hole Interiors

Lower dimensional black holes afford a much more detailed investigation of the
physics
of black hole interiors.  It is reasonable to expect that much more progress
could be
made on incorporating quantum effects due to the mathematical
simplicity of the problem.  Some work has been done in this area
\cite{quantum,Balb},
but there is still much to be done. The possibility of preventing mass
inflation in lower
dimensions \cite{JChan}, raises the intriguing question as to the circumstances
under
which this could occur in higher dimensions. Presumably some form of dilatonic
gravity
will be required.

\end{itemize}

\section*{Acknowledgements}
I am grateful to Jolien Creighton and Jim Chan for their comments to
me during the preparation of this manuscript.
This work was supported by the Natural Sciences and Engineering Research
Council
of Canada.


\begin{thebibliography}{99}

\bibitem{Collas}P. Collas, Am. J. Phys. {\bf 45} (1977) 833
\bibitem{Odintsov}S.D. Odintsov and I.L. Shapiro, Mod. Phys. Lett. {\bf A7}
(1992) 437;
E. Eliadze and S.D. Odintsov, Nucl. Phys. {\bf B399} (1993) 581.
\bibitem{RGRG}R.B. Mann, Gen. Rel. Grav. {\bf 24} (1992) 433.
\bibitem{Harvrev}J. Harvey and A. Strominger, `Quantum Aspects of Black
Holes' hepth-9209055 (EFI-92-41).
\bibitem{cstring}see for example, A. Vilenkin and E.P.S. Shellard,
{\sl Cosmic Strings
and Other Topological Defects}, (Cambridge University Press, 1995).
\bibitem{Polya}A.M. Polyakov, Phys. Lett {\bf B103} (1981) 207.
\bibitem{DesJack}S. Deser, R. Jackiw \& G. 'tHooft, {\sl Ann. Phys.}
	   {\bf 52}, 220(1984);  S. Deser and R. Jackiw, Ann. Phy. {\bf 153} (1984)
405.
\bibitem{Cornish}N. Cornish and Frenkel, Phys. Rev. {\bf D43} (1991) 2555;
Phys. Rev. {\bf D47} (1993) 714.
\bibitem{carlip}S. Carlip in {\sl Proc. of the 5th Canadian Conference on
General Relativity
and Relativisitic Astrophysics}, eds. R.B. Mann and R.G. McLenaghan (World
Scientific, 1994).
 \bibitem{BTZ} M. Banados, C. Teitelboim and J. Zanelli,
  Phys. Rev. Lett. {\bf 69}, 1849 (1992).
\bibitem{rosscoll}R.B. Mann \& S.F. Ross, {\sl Phys. Rev. D} {\bf 47}, 3319
(1993).
\bibitem{rossred}R.B. Mann and S.F. Ross, Class. Quant. Grav. {\bf 10} (1993)
1405.
\bibitem{JackTeit}R. Jackiw, Nucl. Phys. {\bf B252} (1985) 343;
C. Teitelboim, Phys. Lett. {\bf B126} (1983) 41, 46.
\bibitem{MST}R.B. Mann, A. Shiekh, and L. Tarasov, Nucl. Phys. {\bf B341}
(1990) 134.
\bibitem{MFound}R.B. Mann, Found. Phys. Lett. {\bf 4} (1991) 425.
\bibitem{Arnold}A.E. Sikkema and R.B. Mann, Class. Quantum Grav.
{\bf 8} (1991) 219.
\bibitem{Rtstuff}R.B. Mann, S.M. Morsink, A.E. Sikkema and T.G. Steele,
Phys. Rev. {\bf D43} (1991) 3948; S.M. Morsink and R.B. Mann, Class. Quant.
Grav.
{\bf 8} (1991) 2257; J.D. Christensen and R.B. Mann, Class. Quant. Grav. {\bf
9} (1992) 1769;
K.C.K. Chan and R.B. Mann, Class. Quantum Grav.{ \bf{10}}  (1993) 913.
\bibitem{Higher}H.-J. Schmidt, J. Math. Phys. {\bf 32} (1991) 1562.
\bibitem{MSW}G. Mandal, A.M. Sengupta, and S.R. Wadia, Mod. Phys. Lett.
{\bf 6} (1991) 1685; E. Witten, Phys. Rev. {\bf D44} (1991) 314.
\bibitem{DDK}F. David, Mod. Phys. Lett. {\bf A3} (1988) 1651; J. Distler
and H. Kawai, Nucl. Phys. {\bf B321} (1989) 509.
\bibitem{Banks}T. Banks and M. O'Loughlin, Nucl. Phys. B {\bf 362}, (1991) 649
{}.
\bibitem{2dqgm}R.B. Mann, Phys. Lett. {\bf B294} (1992) 310.
\bibitem{MLiou}R.B. Mann, Nucl. Phys. {\bf B418} (1994) 231.
\bibitem{GabJack}J. Gegenberg and G. Kunstatter, gr-qc/9501017.
\bibitem{rtDil}R.B. Mann, Phys. Rev. {\bf D47} (1993) 4438.
\bibitem{Nappi}O. Lechtenfeld and C. Nappi  Phys. Lett. {\bf{B288}} (1992) 72.
\bibitem{Lemos}J.P.S. Lemos \& P.M. Sa, Class. Quant. Grav. {\bf 11}
(1994) L11; Phys. Rev. {\bf D49} (1994) 2897.
\bibitem{Branden}M. Trodden, V. Mukhanov and R. Brandenberger,  Phys. Lett.
{\bf{B316}} 483.
\bibitem{Teo}M.J. Perry and E. Teo  Phys. Rev. Lett.  {\bf{70}} (1993) 2669.
\bibitem{Kchan}K.C.K Chan and R.B. Mann, WATPHYS-TH94/10 gr-qc/9501028 (1995).
\bibitem{Ch3dil}K.C.K. Chan and R.B. Mann, Phys. Rev. {\bf D50} (1994) 6385.
\bibitem{cargeg}S. Carlip. J. Gegenberg and R.B. Mann, gr-qc/9410021 (1994).
\bibitem{HawkEll} S.W. Hawking and G.F.R. Ellis
		    {\em The large scale structure of space-time},
		    New York: Cambridge University Press (1973).
\bibitem{BHTZ}M. Banados, M. Henneaux, C. Teitelboim and J. Zanelli,
	       Phys. Rev. D {\bf 48}, 1506 (1993).
\bibitem{gidd} S. Giddings, J. Abbott and K. Kucha\v{r}, Gen. Rel. Grav.
{\bf 16} (1984) 751.
\bibitem{ross2dcoll}R.B. Mann and S.F. Ross, Class. Quant. Grav.
{\bf 9} (1992) 2335; R.B. Mann, M.S. Morris and S.F. Ross,
Class. Quant. Grav. {\bf 10} (1993) 1477.
\bibitem{BHT}J.D. Brown, M. Henneaux and C. Teitelboim,
Phys. Rev. {\bf D33} (1986) 319; J.D. Brown, {\sl Lower
Dimensional Gravity}, (World Scientific, 1988).
\bibitem{Kriele}M. Kriele, Class. Quant. Grav. {\bf 9} (1992) 1863.
\bibitem{CGHS}C. G. Callan, S. B. Giddings, J. A. Harvey, A. Strominger,
		 Phys. Rev. D {\bf 45}, (1992) 1005; S. Hawking, Phys.
Rev. Lett. {\bf 69} (1992) 406.
\bibitem{Strom}A. Strominger, {\sl Lectures on Black Hole Physics}
hep-th 9501071 (1995).
\bibitem{ADM}R.~Arnowitt, S.~Deser, and C.~W. Misner, in {\sl Gravitation:
An Introduction to Current Research}, edited by L.~Witten (Wiley, New
York, 1962).
\bibitem{York} J.~W. York, {\sl Phys. Rev. D. \bf 33} 2092 (1986);
B.~F. Whiting and J.~W. York, {\sl Phys. Rev. Lett. \bf 61}  1336 (1988);
J.~D. Brown, G.~L. Comer, E.~A. Martinez, J. Melmed,
B.~F. Whiting, and J.~W. York, {\sl Class. Quantum Grav. \bf 7}
1433 (1990); J.~D. Brown, E.~A. Martinez, and J.~W. York, {\sl Phys.
Rev. Lett. \bf 66} 2281 (1991).
\bibitem{Zas}O.~B. Zaslavskii, {\sl Phys. Lett. A. \bf 152} 463 (1991);
O.~B. Zaslavskii, {\sl Class. Quantum Grav. \bf 8} L103 (1991).
\bibitem{BYork} J.~D. Brown and J.~W. York, {\sl Phys. Rev. D. \bf 47} 1407
(1993); J.~D. Brown and J.~W. York, {\sl Phys. Rev. D. \bf 47} 1420
(1993).
\bibitem{Hawkgr} S.~W. Hawking in {\sl General Relativity}, edited by
S.~W. Hawking and W. Israel (Cambridge University Press, Cambridge,
England, 1979).
\item{Wald} R.~M. Wald, in {\sl Black Hole Physics}, edited by
V.~DeSabbata and Z. Zhang (Kluwer Academic Publishers, Dordrecht, 1992).
\bibitem{BCM}J.D. Brown, J. Creighton and R.B. Mann,   Physical Review {\bf
D50}
(1994) 6394.
\bibitem{CLM}D. Cangemi, M. Leblanc and R.B. Mann,    Phys. Rev. D {\bf 48},
3606 (1993).
\bibitem{2djol}J. D. Brown, J. Creighton and R.B. Mann, paper in preparation.
\bibitem{Wald}R. M. Wald, Phys. Rev. D {\bf48} 3427 (1993); R. C. Myers,
 Phys. Rev. {\bf D50} (1994) 6412.
\bibitem{TomRobb}R.B. Mann and T.G. Steele, Class. Quant. Grav. {\bf 9}
(1992) 475.
\bibitem{PCW} P.C.W. Davies, Rep. Prop. Phys. {\bf 41} (1978) 1313.
\bibitem{GP} G.W. Gibbons, M.J. Perry, Proc. R. Soc. Lond. {\bf A358} (1978)
1313.
\bibitem{ptl}P.T. Landsberg and R.B. Mann, Class. Quant. Grav. {\bf 10} (1993)
2373.
\bibitem{Penrose} R. Penrose in {\em Battlle Rencontres}, edited by
		    C.M. De Witt and J.A. Wheeler, New York: Benjamin (1968).
  \bibitem{Poisson} E. Poisson and W. Israel,
		    Phys. Rev. Lett. {\bf 63}, 16, 1663 (1989).
  \bibitem{Ori} A. Ori, Phys. Rev. Lett. {\bf 67}, 7, 789 (1991).
 \bibitem{Bonnano} A. Bonanno, S. Droz, W. Israel and S. Morsink, Physical
Review {\bf D50}
(1994) 7372.
  \bibitem{JChan} J.S.F. Chan and R.B. Mann,
Physical Review {\bf D50} (1994) 7376.
  \bibitem{Droz} S. Droz, Phys. Lett. A {\bf 191}, 211 (1994).
  \bibitem{Husain} V. Husain, Phys. Rev. D {\bf 50}, 2361 (1994).
  \bibitem{CCM} J.S.F. Chan, K.C.K. Chan and R.B. Mann,
		preprint gr-qc/9406049, WATPHYS-TH94/06.
  \bibitem{Jenf} J.S.F. Chan, and R.B. Mann,
		preprint gr-qc/9411064, WATPHYS-TH94/08.
 \bibitem{Price} R.H. Price, Phys. Rev. D {\bf 5}, 10, 2419 (1972).
 \bibitem{Balbinot} R. Balbinot, P.R. Brady, W. Israel and E. Poisson,
		     Phys. Lett. A {\bf 161}, 3, 223 (1991).
  \bibitem{BS} J.S.F. Chan and R.B. Mann,
	       preprint gr-qc/9409034, WATPHYS TH94/07.
\bibitem{quantum} W. Israel in {\em Directions in General Relativity:
		    Proceedings of the 1993 International Symposium, Vol. 1},
		    edited by B.L. Hu, M.P. Ryan Jr. and C.V. Vishveshwara,
		    New York: Cambridge University Press (1993).
  \bibitem{Balb} R. Balbinot and P.R. Brady,
		 Class. Quantum Grav. {\bf 11}, 1763 (1994).

\end{thebibliography}
\end{document}